\documentstyle[12pt]{article}

\begin{document}
%\input pref

%begin macros
\def\nn{\nonumber \\}
\def\be{\begin{equation}}
\def\ee{\end{equation}}
\def\ba{\begin{eqnarray}}
\def\ea{\end{eqnarray}}
\def\la{\label}\def\pl{\plabel}
\def\re{(\ref}

\def\i{{\rm i}}
\let\a=\alpha \let\b=\beta \let\g=\gamma \let\d=\delta
\let\e=\varepsilon \let\ep=\epsilon \let\z=\zeta \let\h=\eta \let\th=\theta
\let\dh=\vartheta \let\k=\kappa \let\l=\lambda \let\m=\mu
\let\n=\nu \let\x=\xi \let\p=\pi \let\r=\rho \let\s=\sigma
\let\t=\tau \let\o=\omega \let\c=\chi \let\ps=\psi
\let\ph=\varphi \let\Ph=\phi \let\PH=\Phi \let\Ps=\Psi
\let\O=\Omega \let\S=\Sigma \let\P=\Pi \let\Th=\Theta
\let\L=\Lambda \let\G=\Gamma \let\D=\Delta

\def\w{\wedge}
\def\0{\over } \def\1{\vec } \def\2{{1\over2}} \def\4{{1\over4}}
\def\5{\bar } %\def\5{\overline }
\def\6{\partial }
\def\7#1{{#1}\llap{/}}
\def\8#1{{\textstyle{#1}}} \def\9#1{{\bf {#1}}}

\def\({\left(} \def\){\right)} \def\<{\langle } \def\>{\rangle }
\def\[{\left[} \def\]{\right]} \def\lb{\left\{} \def\rb{\right\}}
\let\lra=\leftrightarrow \let\LRA=\Leftrightarrow
\let\Ra=\Rightarrow \let\ra=\rightarrow
\def\ul{\underline}

\let\ap=\approx \let\eq=\equiv  %% \let\ex=\times \let\hc=\dagger
        \let\ti=\tilde \let\bl=\biggl \let\br=\biggr
\let\bi=\choose \let\at=\atop \let\mat=\pmatrix
\def\CL{{\cal L}} \def\CD{{\cal D}} \def\rd{{\rm d}} \def\rD{{\rm D}}
\def\CH{{\cal H}} \def\CT{{\cal T}} \def\CM{{\cal M}}
\newcommand{\dR}{\mbox{{\sf I \hspace{-0.8em} R}}}
%end macros

\begin{titlepage}
\renewcommand{\thefootnote}{\fnsymbol{footnote}}
\renewcommand{\baselinestretch}{1.3}
\hfill  TUW - 93 - 26\\
\medskip
\hfill hep-th/9403121\\
\medskip
\vfill

\begin{center}
{\bf {\LARGE{Dirac Quantization of Gravity-Yang-Mills Systems in 1+1
Dimensions}}}
\medskip
\vfill

\renewcommand{\baselinestretch}{1} {\large {
THOMAS STROBL\footnote{e-mail:
tstrobl@email.tuwien.ac.at} \\ \medskip\medskip
\medskip \medskip
Institut f\"ur Theoretische Physik \\
Technische Universit\"at Wien\\
Wiedner Hauptstr. 8-10, A-1040 Vienna\\
Austria\\} }
\end{center}

%\vfill
%\begin{center}
%\end{center}
\vfill
\renewcommand{\baselinestretch}{1}

\begin{abstract}
In two dimensions a large class of gravitational systems including,
e.g., $R^2$-gravity can be quantized exactly also when coupled dynamically
to a Yang-Mills theory. Some previous considerations on the quantization
of pure gravity theories are improved and generalized.
\end{abstract}

\vfill
\hfill Vienna, January 1993  \\
\end{titlepage}

\renewcommand{\baselinestretch}{1}
\setcounter{footnote}{0}
\renewcommand{\thefootnote}{\alph{footnote}}

{\bf 1.} In recent years the study of two-dimensional exactly solvable field
theories has attracted considerable interest. One of the areas of
investigations is 2D Yang-Mills (YM) theory (on a cylinder in a
Hamiltonian approach \cite{Raj}, \cite{Ser}, or  on an arbitrary Riemann
surface when evaluating  the partition function \cite{Mig}); other
models of interest are gravitational ones such as the one for 2D black hole
\cite{CGHS}, the
Jackiw-Teitelboim  model \cite{Jac1},
 or the Katanaev-Volovich  (KV) model \cite{Kat}.
The first main purpose of this letter is to
show that the exact quantum integrability
extends to the combined treatment of the YM theory
and a large class of gravitational systems.\footnote{The classical local
integrability of YM coupled to the KV model, defined through $S_G^{KV}$
below, has been observed already in \cite{Sol}.}

The gravitational part of the action considered in this work will be
\be S_G= \int [\pi_\o  d\o +  \pi_a De^a -  V(\pi_\o,\pi^2) \, \e],
\qquad V = v(\pi_\o) + {\t\0 2} \, \pi^2 , \label{ga} \ee
in which the basic fields are the zweibein and spin-connection
 one-forms $e^a$ and $\o$, respectively, as well as the functions
$\pi_a$ and $\pi_\o$.
$De^a \equiv de^a + \e^a{}_b \o \w e^b$ is
the torsion two-form, $\pi^2 \equiv \pi^a\pi_a \equiv
2\pi_+\pi_-$, and $\e \equiv e^+ \w e^- \equiv e \, d^2x$ with $e\equiv
\det(e_\m{}^a)$ denotes the $\e$-tensor or metric induced
volume-form.
$v$   is some potential and $\t$  a constant.
For the case that
$v$ is chosen as $-(1/4\g)(\pi_\o)^2 + \l$ and
$\t \neq 0$ our action \re{ga})
yields, after elimination of $\pi_a$ and $\pi_\o$ (use $\ast \e =-1$),
2D gravity with torsion \cite{Kat}, \cite{Kum}, \cite{p2}
\be S^{KV}_G = \int [\g d\o \w \ast d\o - {1\0 2\t} De^a \w \ast De_a
- \l \e], \label{KV} \ee
the  most general Lagrangian in two dimensions yielding second order
differential equations  for $e^a$ and $\o$. The same $v$ but with
$\t =0$
is analoguously found to describe torsionless $R^2$ gravity \cite{p5}.
For $V\propto \pi_\o$ the action $S_G$ describes  deSitter
gravity (the Jackiw-Teitelboim model \cite{Jac1}, \cite{many1}),
whereas $V\propto 1/\sqrt{\pi_\o}$
was claimed to effectively describe  4D spherical symmetric gravity
\cite{Kun2}, \cite{Kun}.
$V = const$, furthermore, yields a gravity theory
basically equivalent \cite{Ver} to the string inspired 2D black hole gravity
\cite{CGHS}  for a redefined metric;
this equivalence, however, looses its attractiveness when one couples
the action to nonconformal matter using the redefined metric.
Most of the specific models have been quantized
in a Dirac approach already (cf.\  citations above); moreover,
this is also true  for the general action $S_G$ in the torsionless case
$\t =0$  \cite{Kun}. It is the second main purpose of this letter that
 these quantizations, which came down to the quantization of a
one dimensional point particle system, in many cases
have to be supplemented
 by appropriate discrete indices, originating from
nontrivial topological properties of the constraint surface.

The Yang-Mills part of our action has the standard form
$(1/4\k^2) \int tr (F\w \ast F)$, where $F=dA + A\w A$ and  the trace
is taken in the adjoint representation.
%The coupling to the gravity sector enters via the Hodge star operation.
Rewriting this action  in first order form, it reads
\be S_{YM} = \int tr(EF +  \k^2E^2 \e), \label{YM} \ee
the 'electric fields' $E$ being (Lie algebra valued) functions.
The coupling to the
gravity sector is seen to be separated to the second term now.
For simplicity we will assume the gauge group G to
be compact and simply connected, which implies also that G is simple.
But it would be straightforward to generalize what follows,
e.g., to  arbitrary compact groups G (gaining   a $\Th$-angle for every
U(1)-factor, cf., e.g., \cite{Ser}).

Let us turn to the phase space structure of the theory. Since
$S=S_G+S_{YM}$ is already in first order form, we can  read
off the Poisson bracktes and constraints directly. The canonically
conjugates
are $(e_1{}^a, \o_1, A_1; \pi_a, \pi_\o, E)$, respectively, whereas
the zero components of the basic one forms enforce the following first
class constraints ($\6 \equiv \6 / \6 x^1$)
\ba G_a &=&\6 \pi_a +\e_a{}^b \pi_b \o_1 - \e_{ab} e_1{}^b [V-
\k^2 tr E^2]  \approx 0  \label{Ga}\\
    G_\o &=& \6 \pi_\o + \e^a{}_b \pi_a e_1{}^b \approx 0 \label{Go}\\
    G &=& \nabla_1(A) E \equiv \6_1 E + [A_1,E] \approx 0 \label{G}
\ea
and can be regarded as arbitrary Lagrange multipliers within the
Hamiltonian
\be  H = -\oint dx^1 e_0{}^a G_a + \o_0 G_\o + tr(A_0 G) . \label{H} \ee
We observe  that in two dimensions the addition of a dynamical
gravity sector leaves the Yang-Mills' Gau{\ss} law $G\approx 0$
completely unchanged. This is in contrast to four space time dimensions
where the covariant derivative $\nabla$ containes also a gravitational
connection resulting from the fact that the electric fields  are not
functions there but one forms on a three manifold (c.f., e.g., \cite{Ash}).
Since, furthermore, $\6 (tr E^2) = tr G E/2 \approx 0$, on-shell the
YM theory modifies the gravitational theory only via  dynamically
shifting the cosmological constant of the gravity sector by the
YM Hamiltonian $H^{(0)}_{YM} \equiv -
\k^2 \oint tr E^2 dx^1$.\footnote{This suggests
also that there should be some connection to \cite{Jac2} where
the authors
allowed for a dynamical cosmological constant within the reformulated
2D black hole gravity so as to reinterpret the resulting theory as
a connection flat gauge theory for the centrally extended Poincar\'e
group. Indeed, choosing as our potential $V =1/4\k^2$,  as
YM gauge group the real line, and shifting $E$ by $-1/2\k^2$, the limit
$\k\to 0$ reproduces that theory.}

{\bf 2.} To quantize the system we  choose our wave functionals to depend
on $\pi_a, \pi_\o, A_1$ and replace $e_1{}^a,\o_1,E$ by the appropriate
derivative operators. Our ordering prescription for the quantum
constraints is to put all derivative operators to the right, which
yields a consistent quantum algebra.  For the
solution of the quantum Gau{\ss} law  $G \Psi =0$ we can refer  the reader
to
the extensive literature. The basic result is (cf., e.g., \cite{Ser})
that the functional  $\Psi[\pi_b,\pi_\o,A_1(x^1)]$ can be written as
a function $\Psi[\pi_b,\pi_\o,a]$ of a constant
 element $a$ of the corresponding Cartan subalgebra (CSA) which is  gauge
related to $A_1(x^1)$; due
to a residual gauge freedom,  $\Psi$ is, moreover,  invariant under
affine Weyl transformations so that  the fundamental domain of definition
of
$\Psi$ as a function of $a$ is the Weyl cell of the CSA.
For the simplest case
of $SU(2)$, e.g., $\Psi$  is a periodic function of $a \in R$,
 and similarily for the case of $SU(3)$ a function on a triangle,
'periodically' continued to the plane via Weyl reflections and
translations.
The Hamilton operator of pure Yang-Mills theory $H^{(0)}_{YM}$ projected
onto this physical subspace is proportional to the ordinary Laplacian on
the
CSA. The same projection yields also a natural measure $\m(a)$
with respect to which  $H^{(0)}_{YM}$ is  self-adjoint.

Due to the finite
size of the Weyl cell $H^{(0)}_{YM}$ has a discrete spectrum $\epsilon_k$,
$k \in N$,
and  we can expand our gravity-Yang-Mills functional
$\Psi$ onto an orthogonal set of eigenfunctions
$\chi_{k,l}(a)$,  $l$ labelling possible
degeneracies of $\ep_k$:
\be \Psi[\pi_+,\pi_-,\pi_\o,a] = \sum_{k,l} \psi_{k,l}[\pi_+,\pi_-,\pi_\o]
\, \chi_{k,l}(a) . \label{exp} \ee

Applying next the remaining quantum constraints $G_\pm, G_\o$ to this
expansion, we see that {\em each} of the $\psi_{k,l}$ has to be annihilated
by the corresponding  operators in which $-\k^2  tr E^2(x^1)$ has been
replaced by $\ep_k$. Let us denote these modified operators by
$G_{\pm,(k)}$
and analoguosly $v_{(k)}:=v + \ep_k$,
$V_{(k)}:=V + \ep_k  \equiv v_{(k)}(\pi_\o) + \pi^2/2$.
Next one finds the combination
\be \pi^a G_{a,(k)} + V_{(k)} G_\o = \2 \6 (\pi^2) + V_{(k)} \6 \pi_\o
\la{d} \ee
to act in a purely multiplicative way on the wave functionals.
Multiplying \re{d})
from the left  by the integrating factor $\exp(\t \pi_\o)$, this yields
(no sums)
\ba &\6 Q_{(k)} \psi_{k,l} = 0 & \la{dq} \\
&Q_{(k)}(\pi^2,\pi_\o) = \2 \pi^2 \exp(\t \pi_\o) +
\int^{\pi_\o}  v_{(k)}(u) \exp(\t u) \, du.& \la{q} \ea
This is a restriction to the support or the domain of definition of
$\psi_{k,l}$.
Further inspection of the constraints  show
that the only 'allowed'
functions $\pi_a(x^1)$, $\pi_\o(x^1)$
containing the   'critical points'
\be \pi_+=\pi_-=0, \pi_\o =\a_c, \qquad \a_c: V_{(k)}(\a_c)=0 \la{crit} \ee
are the constant ones.
In this restricted domain of definition (!) the
wave functional $\exp(\phi[\pi_+,\pi_-,\pi_\o])$ with
\be \phi = \left\{
\begin{array}{l} \qquad \qquad 0 \qquad \qquad
\qquad \qquad \qquad \qquad \qquad \! : \pi_+\equiv \pi_- \equiv 0 \\
 -{i\0 \hbar} \oint ln \mid \pi_+ \mid d\pi_\o =
{i\0 \hbar} \oint ln \mid \pi_- \mid d\pi_\o \qquad : \mbox{otherwise}
  \end{array} \right. \la{phase} \ee
can be  seen to be a particular (continuous) solution to the quantum
constraints. Thus the general solution to the gravity constraints
can be written as
\be \psi_{k,l} = \exp(\phi) \,\, \ti \psi_{k,l} \la{fac} \ee
with a $\ti \psi_{k,l}$ which is invariant under the Lie derivative part
of the $G_{a,(k)},G_\o$ constraints, i.e. under infinitesimal classical
gauge transformations. At this point  it is
helpful to interpret $\ti \psi_{k,l}$ (or also $\psi_{k,l}$)
as a  functional of
(parametrized and connected) loops in the three dimensional
target space $(\pi_+,\pi_-,\pi_\o)$, or better on the two-surfaces $\CM_q$
generated by setting $Q_{(k)}$ to a (varying)  constant $q$
(cf.\ Eq.\ \re{dq})).
Now, on any of these surfaces the flow of the constraints
is transitive, except precisely for the critical points \re{crit}),
which are fixed points under this flow. This is most easily seen by noting
that on any connected part of this surface where $\pi_+ \neq 0$ [resp.
$\pi_-
\neq 0$] we can use $(G_\o/\pi_+, G_+/\pi_+)$ [$(G_\o/\pi_-, G_-/\pi_-)$]
as conjugate variables to the local coordinates
$(\pi_+,\pi_\o)$ [$(\pi_-,\pi_\o)$] of $\CM_q$; furthermore $\{\pi_a,G_b\}
=- \e_{ab} V \d$. However, the wavefunctions
$\ti \psi_{k,l}$ do not depend  only on $q$ as one might suppose at first
sight. This is so because certainly only loops from the same homotopy class
and the same
component of $\CM_q$ can be deformed into each other by means of the
constraints. Thus $\ti \psi_{k,l}$ is a function of $q$, $\pi_0(\CM_q)$,
and
$\pi_1(\CM_q)$ (as well as the fixed points, if $q=q_c \equiv
Q_{(k)}(0,\a_c)$).
Labelling the elements of the  latter two discrete groups by
$n^{q}_0$ and $n^{q}_1$, and suppressing the fixed points for a moment,
we find  \be \ti \psi_{k,l} = \ti \psi_{k,l}(q,n^{q}_0,n^{q}_1),
\la{wa} \ee
which together with (\ref{exp}, \ref{phase}, \ref{fac}) describes the
general solution of the quantum constraints (\ref{Ga} - \ref{G}).

To illustrate  the above  considerations, let us regard
some examples: $V_{(k)} = \pi_\o$ (for some fixed $k$,
dropping this index furtheron within the paragraph) implies
$Q= \2 [\pi^2 + (\pi_\o)^2]$. Putting this to a constant $q$,
we  obtain the typical Lorentz orbits in a three dimensional
'Minkowski space'\footnote{This has to do with the fact that
the gravitational action for the above potential can be reinterpreted
as the one of a $\pi F$-theory for gauge group $SO(2,1)_e \sim PSL(2,R)$
\cite{many1}, or rather its universal covering,
as pointed out in \cite{p6}, \cite{p8}.}, i.e. two-sheet hyperboloids
for  $q  >0$, which implies  $n^+_0 \in \{1,2\}$ and   $n^+_1=0$,
one-sheet hyperboloids for $q<0$, which implies $n^-_0=0$ and
$n^-_1 \in Z$,
as well as the future and past lightcone 'separated' by the
origin for the critical value $q=q_c=0$.
Due to the latter parts the resulting
orbit space is non-Hausdorff (such as the Lorentz orbit space),
and there arises some  arbitrariness in determining $n^0_{0,1}$ for this
value of $q$:
The origin and the  light cones have no disjoint
neighborhoods  so that, e.g., continuous
functions on this space would identify them ($\Ra n^0_0=0$, $n^0_1 =0$);
on the other
hand we know that there are no loops passing through the origin since
it is critical ($\Ra n^0_0 \in \{1,2\}$, $n^0_1 \in Z$, plus the
origin as an own orbit).

As a second example let us consider $R^2$-gravity with potential
$V=3\pi_\o^2/2-3/ 4$ coupled to $SU(2)$-YM, yielding
$\ep_k = k^2/4$ (for $\k^2 = 2$, if we choose $x^1=0 \sim x^1=1$).
We obtain  $2Q_{(k)} =\pi^2+\pi_\o^3 +(2\ep_k-3/2)\pi_\o$.
For $k=1$ there are two critical values of $q$:
$\a_c = \pm 1/\sqrt{3} \ra q_c  = \mp 1/\sqrt{27}$, whereas
for $k =2,3,...$ there are no critical values of $q$. In the latter case
the resulting two-surfaces are all connected and simply connected.
This is also true for $\ep = \ep_1 =1/4$ and  $q \not\in \,
[-1/\sqrt{27},
1/\sqrt{27}]$, whereas in the case $k=0$ and $q \in \, ]-1/\sqrt{27},
1/\sqrt{27}[\,$ the $Q_0 =q$ surface is connected but has the fundamental
group of a pointed torus. At $q=q_c$ there arises a similar situation
as in the first example. Thus in this example the
wave functions have the form
\be \Psi = \exp(\phi) [\ti \psi_1(q,n_1) \chi_1(a) + \sum_{k \ge 2}
\ti \psi_k(q) \chi_k(a)]  \ee
where $\phi$ is defined in \re{phase}), $\chi_k$ is a periodic
function of one argument, and $\psi_k$ is a function of one unbounded
variable except for $k=1,\,n_1 \neq 0$, in which case it has support
$[-1/\sqrt{27},1/\sqrt{27}]$.

More generally the
situation can be depicted as follows: For any fixed value
$q$ and $\ep_k$ equation \re{q}) induces a curve in a $\pi^2$ over
$\pi_\o$ diagram. If this curve has no intersections with the $\pi_\o$
axis, $\CM_q$ has two simply connected components. Otherwise (and
for $q\neq q_c$) $\CM_q$ is always
connected and the number of basic non-contractible loops is by one larger
than the number of intersections  of the $\pi^2(\pi_\o)$-curve with the
$\pi_\o$ axis.  A change of this number  can occur only at
critical values of $q$, which correspond to curves having at
least one sliding intersection with the
$\pi_\o$ axis. All these surfaces  $\CM_q$ are noncompact and the
spectrum of $q$ ranges over all of $\dR$.\footnote{This picture
is changed
when regarding the gravitational theories  corresponding to
a Euclidean signature.
Some values of $q$ generate compact target-space surfaces $\CM_q$
then; on the latter
the Wick rotated phase factor \re{phase}) is globally defined only
for some values of $q$, which leads to a discretization of the
corresponding part of the spectrum, \cite{p8}, \cite{p9},
\cite{Ama}.}

{\bf 3.} Let us conclude with some remarks.  Firstly, already the example of
Lorentz transformations in a Minkowski space
shows that in general orbits cannot be (uniquely) characterized by means of
continuous invariants only: Beside the invariant length of a vector of
this space, one needs also some (discontinuous) sign functions and, to
distinguish the origin from the light cones, even a distribution.
This is the reason for the quantum numbers $n_0$ and $n_1$
within the wave functions $\psi_{k,l}$: $Q_{(k)}$
is the  contiunous invariant on the underlying function space,
$n_0$ and $n_1$ correspond to discontinuos (invariant) functions on
the latter,  and \re{crit}) is the counterpart
to the origin of the Minkowski space example above.
{}From this perspective it comes  at no surprise that beside \re{wa})  also
\be \sum_c C_c \d[\pi_+]\d[\pi_-]\d[\pi_\o(x^1) -\a_c], \quad C_c \in
{\bf C}
\la{dis} \ee solve the quantum constraints.

Secondly, we still have to define an inner product for the
$\ti \psi_{k,l}$ (for fixed $k$ and $l$ only, since the $\chi_{k,l}$
in \re{exp}) are orthogonal by construction).
On parts of the phase space which do not contain critical points \re{crit})
the  Dirac observable  conjugate to $q = \oint Q_{(k)} dx^1$ can be put
into the form
\be p = \oint \exp(-\t \pi_\o) {e_1{}^-\0\pi_+}dx^1 \sim
\oint \exp(-\t \pi_\o) {e_1{}^+\0\pi_-} dx^1. \ee
Replacing $e_1{}^\pm$ by the corresponding functional derivative
operator, it acts as $(\hbar/i) (d/dq)$ on $\ti \psi_{k,l}$.
Requiring that this fundamental Dirac observable shall be represented by a
hermitean operator, restricts the measure to be proportional
to $dq$ within any interval of $q$ not containing a critical value $q_c$.
The implementation
of the quantum numbers $n_0$ and $n_1$, however, seems not determined
by this procedure and a further investigation of this point would
be interesting.

Let me further remark that it is probably incorrect
to just neglect the solutions \re{dis}). This becomes most
apparent in the extreme case $V_{(k)} \equiv 0$, where
any constant loop  on the  $\pi_\o$ axis becomes critical;
on the  classical level these solutions are pared
with the  compactifications of Minkowski space along the  boost orbits,
yielding Misners two dimensional analogue of a Taub-Nut space
\cite{Mis}. So, neglecting the solutions
\re{dis}) in this case comes down to
throwing away about a third of the reduced phase space.
(The remaining 'two
thirds', represented by \re{wa}) on the quantum level,
 correspond to a Minkowski space factored along the translational
isometries of the flat metric, thus, in part,
also incorporate classical
solutions with closed timelike curves). In the more generic case
$V_{(k)} \not\equiv 0$ the neglection of \re{dis}) would still change
the degeneracy of the spectrum of the Dirac observable $q$.
(In the Euclidean
formulation of the theory the omission of \re{dis}) may
even lead to a change of the spectrum of $q$ \cite{p8}).

In this letter we carefully constructed the general solutions to
the quantum constraints of many gravity theories coupled to YM.
Open technical questions concern the construction of an inner
product and a possible inclusion of the solutions \re{dis}).
Furthermore, the treatment of conceptual questions of quantum
gravity seems rewarding at this point: firstly, since the classical
solutions
include black hole type solutions for some choices of $V$, and,
secondly, since  $S$ reduces to a reparametrization invariant
formulation of a pure 2D YM theory for $V \equiv 0$ so
that the models comprised in the action (\ref{ga}, \ref{YM})
 may well be used for testing and developing
concepts to solve the 'problem of time' \cite{Ish} (cf.\ pcite{p2},
\cite{p8}). From the mathematical
point of view the evaluation of the partition function for $S_G$
would be an interesting open task (cf.\ also \cite{Kaw}).
Also, $S_G$ can be regarded as a generalization of an $SO(2,1)$ gauge
theory; proceeding similarily for other groups or target space
dimensions one obtains further new topological 2D field theories
\cite{p9}.

\vskip5mm

As always I am grateful to P.\ Schaller for valuable discussions.

\vspace{- 10 pt}
\renewcommand{\Large}{\normalsize}

\end{document}